%% file: main.tex
\documentclass[conference]{IEEEtran}
\IEEEoverridecommandlockouts
\usepackage{cite}
\usepackage{amsmath,amssymb,amsfonts}
\usepackage{amsthm}
\usepackage{algorithmic}
\usepackage{graphicx}
\usepackage{textcomp}
\usepackage{xcolor}
\usepackage{tikz}
\usepackage{setspace}
\usepackage{mathtools}
\usepackage{color}
\usepackage{caption}
\usepackage{subcaption}
\usepackage{lipsum}
\usepackage{soul}
\usepackage{pgfplots}
\usetikzlibrary{automata, positioning, arrows}
\usepackage{tabularx, booktabs}
\newcolumntype{C}{>{\centering\arraybackslash}X} 
\usepackage{wrapfig}
\usepackage{ragged2e}
\usepackage{lineno,hyperref}
\usepackage{float}
 
\graphicspath{ {images/} }
\definecolor{ao}{rgb}{0.0, 0.5, 0.0}
\usepackage{graphicx}
\modulolinenumbers[5]
\usepackage{adjustbox}
\newtheorem{definition}{Definition}
\def\BibTeX{{\rm B\kern-.05em{\sc i\kern-.025em b}\kern-.08em
    T\kern-.1667em\lower.7ex\hbox{E}\kern-.125emX}}
\usepackage[linesnumbered,ruled,vlined]{algorithm2e}
\usepackage{multirow}
\usepackage{epstopdf}
\usepackage{url}
\usepackage{soul,color}
\usepackage{booktabs}
\usepackage{colortbl}
\usepackage[T1]{fontenc}
\usepackage{multicol}
\usepackage{amsmath}
\usetikzlibrary{patterns}

\newcommand{\superscript}[1]{\ensuremath{^{\textrm{#1}}}}
\usetikzlibrary{shapes.gates.logic.US,trees,positioning,arrows,topaths}
\begin{document}
\title{Attack Trees for Security and Privacy in Social \\ Virtual Reality Learning Environments}

\author{Samaikya~Valluripally,
Aniket~Gulhane,
Reshmi Mitra\textsuperscript{*}, Khaza Anuarul Hoque,
Prasad Calyam\\

University of Missouri-Columbia, \textsuperscript{*}Webster University \\

\small{\{svbqb, arggm8\}@mail.missouri.edu, \{calyamp, hoquek\}@missouri.edu}, reshmimitra25@webster.edu

\thanks{This material is based upon work supported by the National Science Foundation under Award Number CNS-1647213. Any opinions, findings, and conclusions or recommendations expressed in this publication are those of the authors and do not necessarily reflect the views of the National Science Foundation.}
}

\maketitle
\begin{abstract}
Social Virtual Reality Learning Environment (VRLE) is a novel edge computing platform for collaboration amongst distributed users. Given that VRLEs are used for critical applications (e.g., special education, public safety training), it is important to ensure security and privacy issues. In this paper, we present a novel framework to obtain quantitative assessments of threats and vulnerabilities for VRLEs. Based on the use cases from an actual social VRLE viz., vSocial, we first model the security and privacy using the attack trees. Subsequently, these attack trees are converted into stochastic timed automata representations that allow for rigorous statistical model checking. Such an analysis helps us adopt pertinent design principles such as \textit{hardening}, \textit{diversity} and \textit{principle of least privilege} to enhance the resilience of social VRLEs.  Through experiments in a vSocial case study, we demonstrate the effectiveness of our attack tree modeling with a reduction of 26\% in probability of loss of integrity (security) and 80\% in privacy leakage (privacy) in before and after scenarios pertaining to the adoption of the design principles.
\end{abstract}

\section{Introduction}
\label{Sec:Intro}
Social Virtual Reality (VR) is a new paradigm of collaboration systems that uses edge computing for novel application areas involving virtual reality learning environments (VRLE) for special education, surgical training, and flight simulators. Typical VR system applications comprise of interactions that require coordination of diverse user actions from multiple Internet-of-Things (IoT) devices, processing activity data and projecting visualization to achieve cooperative tasks. However, this flexibility necessitates seamless interactions with IoT devices, geographically distributed users outside the system's safe boundary, which poses serious threats to security and privacy~\cite{Educause}. 

Although existing works~\cite{IoT_Security_Privacy, SPS_Accelerated, new-haven} highlight the importance of security and privacy issues in VR applications, there are a limited systematic efforts in evaluating the effect of various threat scenarios on such edge computing based collaborative systems with IoT devices. Specifically, VRLE applications are highly susceptible to Distributed Denial of Service (DDoS) attacks, due to the distributed IoT devices (i.e.,VR headsets) connecting to virtual classrooms through custom controlled collaboration settings. Moreover, loss of confidential information is possible as VRLEs track student engagement and other realtime session metadata. 

\begin{figure}
  \centering
  \includegraphics[width=\linewidth]{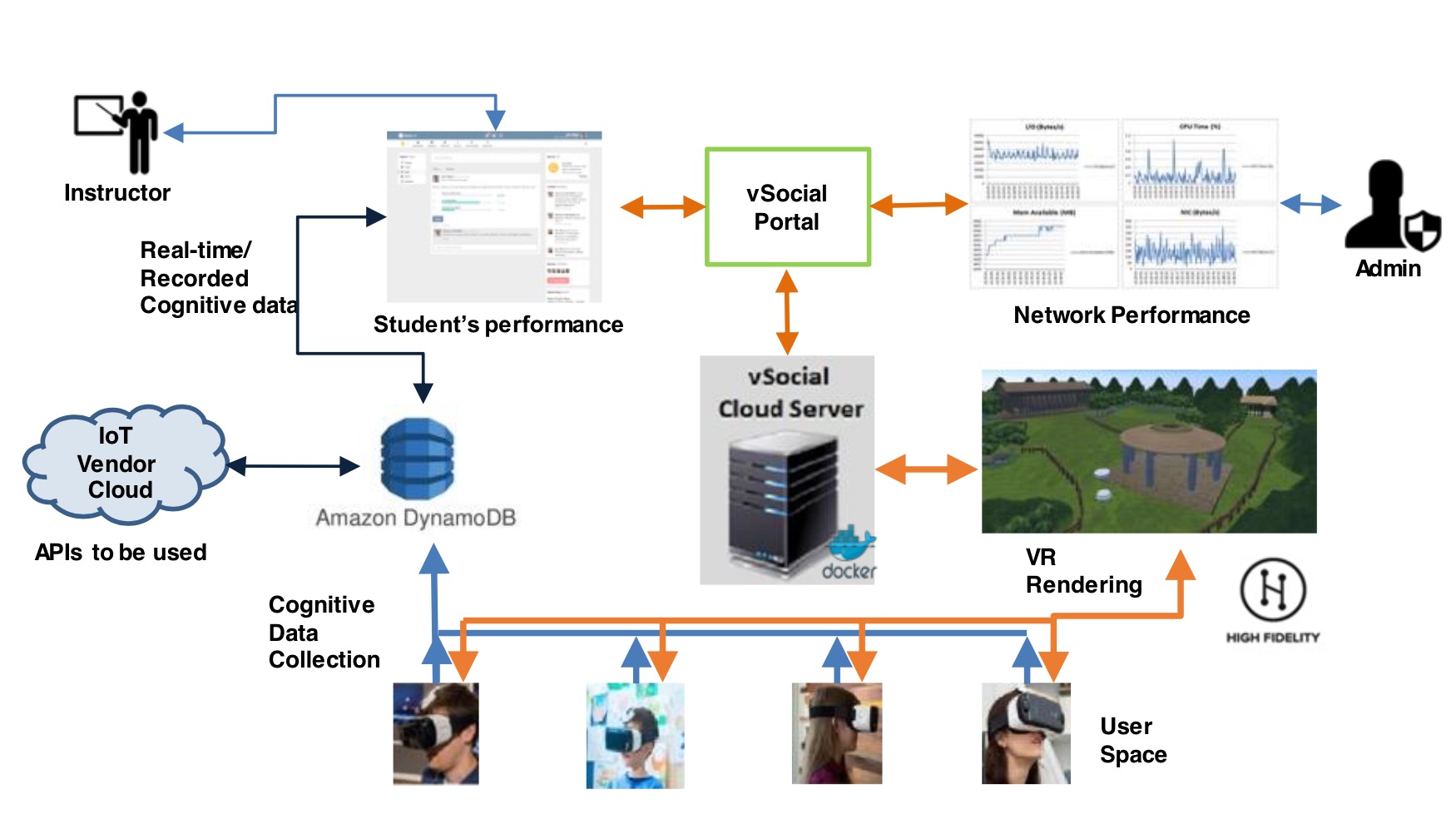}
  \captionsetup{justification=centering}
  \caption{\footnotesize{vSocial system components used for real-time student learning session management.}}
  \label{fig:vSocial_Arch}
  \vspace{-6mm}
\end{figure}

In this paper, we consider a VRLE application designed for youth with autism spectrum disorder (ASD) as case study viz., vSocial~\cite{vSocial}\footnote{Moving forward we use the acronym VRLE to sometimes interchangeably refer to our case study application viz., vSocial.}. 
Our multi-modal VRLE system as shown in Figure~\ref{fig:vSocial_Arch} uses the High Fidelity platform~\cite{High-Fidelity}, and renders 3D visualizations based on the dynamic human computer interactions with an edge cloud i.e., vSocial Cloud Server. VRLE setup includes: VR headset devices (HTC Vive), hand-held controllers, and base stations for accurate localization and tracking of controllers~\cite{vSocial}. Any disruption caused by an attacker with malicious intent on the instructor's VR content or administrator privileges will impact user activities and even cause cybersickness. Failure to address these security and privacy issues may result in alteration of instructional content, compromise of learning outcomes, access privileges leading to confidential student information disclosure and/or poor student engagement in ongoing classroom sessions.
Motivated by the importance of ensuring security and privacy in a VRLE application, we propose a novel framework for quantitative evaluation of security and privacy metrics inspired by the approach discussed in \cite{AFT}. Our proposed framework has benefits for identifying potential security and privacy attacks caused by vulnerabilities in a VRLE application in a manner that is not possible with traditional analysis~\cite{TR1}. We model the potential security and privacy threats using attack tree formalism and then convert them into their equivalent stochastic timed automata (STA) representations~\cite{STA_Advantages}. The STAs are then analyzed using the statistical model checking technique (SMC)~\cite{UPPAAL SMC Tutorial}. The SMC technique is widely used owing to its capability of modeling and analyzing complex stochastic and dynamic system behaviors~\cite{Souri2019}, and thus can be used to formally verify the VRLE system and user behaviors.

We use the attack trees concept from~\cite{ATSH} and derive graphical models that provide a systematic representation of various attack scenarios. Although attack trees are popular, they lack support for modeling the temporal dependencies between the attack tree components. To overcome this limitation, we utilize an automated state-of-the-art SMC tool UPPAAL~\cite{UPPAAL SMC Tutorial}. Our approach overall involves translating the constructed security and privacy attack tree of the VRLE application into the Stochastic Timed Automata (STA) in a compositional manner. For the purposes of this paper, we define: (a) \textit{security} -- as a condition that ensures a VR system to perform critical functions with the establishment of confidentiality, integrity, and availability~\cite{NIST}, and (b) \textit{privacy} -- as a property that regulates the IoT data collection, protection, and secrecy in interactive systems~\cite{NIST}.


The main paper contributions summary is as follows:\\
-- We propose a framework to evaluate security and privacy of VR applications using the attack tree formalism and statistical model checking. To show the effectiveness of our solution approach, we utilize a VRLE application case study viz., vSocial that uses edge computing assisted IoT devices for students and instructor(s) collaboration.\\ 
-- We perform a trade-off analysis by evaluating the severity of different types of attacks on the considered VRLE application. From our results, we observe that: i) unauthorized access and causes of DoS attack (in security attack tree), ii) track user movement and user physical location (in privacy attack tree) are the most vulnerable candidates in a VRLE.\\
-- We demonstrate the effectiveness of using design principles (also known as security principles) i.e., \textit{hardening, diversity, principle of least privilege} on the privacy and security of VRLE applications in the event of most severe threats. We show that in terms of security {\textbf{--}} a combination of \{\textit{hardening, principle of least privilege}\} is most influential in reducing the probability of Loss of Integrity (LoI). Similarly for privacy {\textbf{--}} a combination of \{\textit{diversity, principle of least privilege}\} is most influential in reducing privacy leakage.

The remainder of the paper is organized as follows: Section~\ref{Sec:relwork} discusses related works. Section~\ref{Sec:Preliminaries} introduces the necessary background and terminology. Section~\ref{Sec:Problem} discusses the proposed security and privacy framework in detail. Section~\ref{Sec:experiment} presents the numerical results using our proposed framework on the VRLE case study. Section~\ref{Sec:Design-principles} discusses the effectiveness of design principles on the security and privacy threat scenarios. Section~\ref{Sec:conclusion} concludes the paper. 

\section{Related Works}
\label{Sec:relwork}
There have been several comprehensive studies that highlight the importance of security and privacy threats on IoT and related paradigms such as Augmented Reality (AR) with IoT devices, and edge computing. A recent study~\cite{Educause} on challenges in AR and VR discusses the threat vectors for educational initiatives without characterizing the attack impact. Survey articles~\cite{IoT_Security_Privacy,SPS_Accelerated,new-haven,Threat_model_fog_computing,privacy_fog_computing,IoT_security} are significant for understanding the concepts of threat taxonomy and attack surface area of IoT and fog computing. They highlight the need to go beyond specific components such as network, hardware or application, and propose end-to-end solutions that consider system and data vulnerabilities. An observation given the above state-of-art is that there are very few scholarly works on the quantitative evaluation for these security and privacy threats in the context of VR applications. 

We borrow the attack trees concept that is used commonly in cyber-physical systems involving SCADA system~\cite{SCADA}, and adapt it for threat modeling to determine the probability of detection and severity of threats. One of our preliminary works~\cite{CCNC} showed risk assessment of security, privacy and safety metrics of the VRLE applications utilizing an attack tree simulation tool. In contrast, this work focus is on formal modeling of attack trees using STAs and utilizes a state-of-the-art formal verification tool to evaluate the developed security and privacy attack trees. In addition, our proposed framework incorporates the concept of design principles to enhance the security and privacy of VRLE applications.

Amongst the numerous prior works on attack trees, the work in~\cite{AFT} presents a novel concept of Attack Fault Tree (AFT), a combination of both attack and fault trees. A model of STA~\cite{STA_Advantages} alleviates some assumptions made in timed automata and provides advantages such as choice of transitions requiring satisfaction of very precise clock constraints. Timed automata~\cite{NSTA} provides an abstract model of the real system by using clocks as well as timing constraints on the transition edges. As compared to Continuous-Time Markov Chains (CTMC)~\cite{CTMC}, STA models allow us to express hard time constraints such as \textit{x}$\leq$\textit{t}. We studied the above existing modeling techniques to formalize our security and privacy attack trees into STA, which we evaluate using a model checking tool viz., the UPPAAL SMC~\cite{UPPAAL}.

\section{Background and Terminology}
\label{Sec:Preliminaries}
\subsection{Attacks in VRLE application use case:}
The users in a social VRLE are networked and geographically distributed, which creates a series of potential attack scenarios. Using vSocial shown in Figure~\ref{fig:vSocial_Arch} as a social VRLE case study, herein we demonstrate exemplar security and privacy attacks that can affect the VRLE application sessions. 

\begin{figure}
  \centering
  \includegraphics[width=0.5\textwidth]{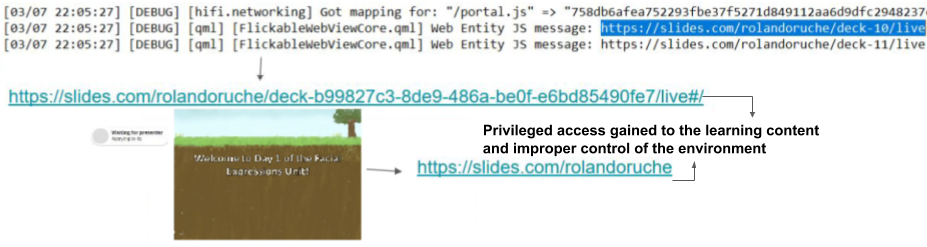}
  \captionsetup{justification=centering}
  \caption{\footnotesize{Unauthorized access to VRLE learning content.}}
  \label{data}
  \vspace{-4mm}
\end{figure}

\noindent \textit{\bf{Security Attacks:}} An attacker can gain unauthorized access to either tamper any confidential information (user, network, VRLE components) by impersonating as a valid user, or disclose compromised confidential information. To elucidate, the instructional content in a vSocial application is in a web-enabled presentation format using the features present in High Fidelity. To guide the students through activities in the vSocial learning environment, the instructor will have privileged access to control the learning content settings such as e.g., editing the slides, and rewarding the students based on their performance. Gaining \textit{Unauthorized access} to the instructor account as shown in Figure~\ref{data} can lead to \textit{disclosure of user information}, and tampering of the learning content in vSocial to negatively impact the users' (students') learning experience. 
\begin{figure}[h]
  \centering
  \vspace{-3mm}
  \includegraphics[width=\linewidth]{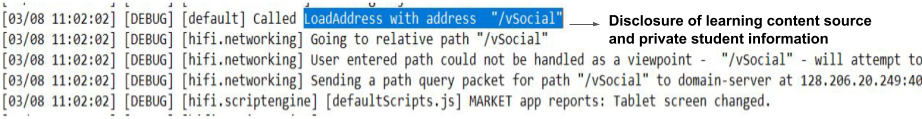}
  \captionsetup{justification=centering}
  \caption{\footnotesize{Privacy attack on vSocial application.}}
  \label{VRSpace}
 \end{figure}
\\
\noindent \textit{\bf{Privacy Attacks:}}
A user privacy breach can involve an intruder entering a VRLE world with fake credentials to snoop into the virtual classroom conversations. The attacker can then disrupt an ongoing VRLE session by obstructing the view of the users in their learning sessions and can even disorient the content. Disorientation can possibly lead to a user running into a wall and getting physically hurt. Privacy attacks can also involve \textit{packet tampering} that was demonstrated in \cite{CCNC}, where an attacker performs illegal packet capture in order to extract sensitive information (\textit{packet sniffing attack}). A potential privacy breach can occur when the attacker \textit{discloses the confidential information} obtained from the captured packets as shown in Figure~\ref{VRSpace}. Using this packet sniffing attack, the avatar (user virtual information) can also be disclosed with private location information and student credentials.
\subsection{Statistical model checking} 
\label{STA-ex}
Statistical model checking (SMC) is a variation of the well-known classical model checking~\cite{model-checking-book} approach for a system that exhibits stochastic behavior. The SMC approach to solve the model checking problem involves simulating (Monte Carlo simulation) the system for finitely many runs, and using hypothesis testing to infer whether the samples provide a statistical evidence for the satisfaction or violation of the specification~\cite{PMC}.\\
\noindent \textbf{Stochastic timed automata:} Stochastic timed automata (STA) is an extended version of timed automata (TA) with stochastic semantics. A STA associates logical locations with continuous, generally distributed sojourn times~\cite{NSTA}. In STA, constraints on edges and invariants on locations, such as clocks are used to enable transition from one state to another~\cite{AFT}. 
\vspace{-2mm}
\begin{definition}[\hspace{-2mm} Stochastic timed automata]
\label{STA_Formaldefinition}
Given a timed automata which is equipped with assignment of invariants $\mathcal{I}$ to locations $\mathcal{L}$, we formulate an STA as 
a tuple T = $\langle$ $\mathcal{L}$, l\textsubscript{init}, $\Sigma$, $\mathcal{X}$, $\mathcal{E}$, $\mathcal{I}$,  $\mu$ $\rangle$,
where $\mathcal{L}$ is a finite set of locations,
l\textsubscript{init} $\in$ $\mathcal{L}$ is the initial location,
$\Sigma$ is a finite set of actions,
$\mathcal{X}$ is the finite set of clocks,
$\mathcal{E}$ $\subseteq$ $\mathcal{L}$ $\times$ $\mathcal{L\textsubscript{clk}}$ $\times$ $\Sigma$ $\times$ $2^\mathcal{X}$ is a finite set of edges,
with $\mathcal{L\textsubscript{clk}}$ representing the set of clock constraints,
$\mathcal{I}$: $\mathcal{L}$ $\longrightarrow$ $\lambda$ is the invariant where $\lambda$ is the rate of exponential assigned to the locations $\mathcal{L}$,
$\mu$ is the probability density function ( $\mu$\textsubscript{l}) at a location $l \in \mathcal{L}$.
\end{definition}
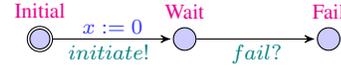
\begin{figure}[h]
\vspace{-4mm}
\centering
\small
\input{STA_Tikz/Top_Event.tex}
\caption{\footnotesize{An exemplar STA.}}
\label{fig:Sample_STA}
\vspace{-4.5mm}
\end{figure}
An exemplar STA is shown in Figure~\ref{fig:Sample_STA} that consists of the locations \{Initial, Wait, Fail\}. Herein, the $Initial$ location represents the start of execution of an STA and a $clock$ $x$ is used to keep track of the global time. The communication in an STA exists between its components using message broadcast signals in a bottom-up approach. The STA is activated by broadcasting \textit{initiate!} signal, which transitions to wait location and waits for the \textit{fail} signal. 
\setlength\abovedisplayskip{0pt}
In an STA, time delays are governed as probability distributions (used as invariants) over the locations. The Network of Stochastic Timed Automata (NSTA) is defined by composing all component automaton to obtain a complete stochastic system satisfying the general compositionality criterion of TA transition rules~\cite{NSTA,UPPAAL SMC Tutorial}.\\
\noindent \textbf{UPPAAL SMC:} UPPAAL SMC is an integrated tool for modeling, validation, and verification of real-time systems modeled as a network of stochastic timed automata (NSTA) extended with integer variable, invariant, and channel synchronizations~\cite{UPPAAL}. In SMC, the probability estimate is derived using an estimation algorithm and statistical parameters, such as $1-\alpha$ (required confidence interval) and $\epsilon$ (error bound)~\cite{SMC_Semantics}. For instance, if we indicate goal state in the STA of $Top\_event$ as $Fail$, then the probability of a successful occurrence within time $t$ can be written as: $Pr[x <= t] (<> Top\_event.Fail)$
where, $<>$ represents the existential operator ($\Diamond$) and $x$ is a clock in the STA to track the global time.

\subsection{Design principles}
\label{definition-design-principles}
\noindent 
To build a trustworthy VRLE system architecture which ensures security and privacy, integration of design principles in the life cycle of edge computing interconnected and distributed IoT device based systems is essential~\cite{Designprinciples}. We adapt the following three design principles from NIST SP800-160~\cite{NIST,Designprinciples} such as: (i) \textit{Hardening} -- defined as reinforcement of individual or types of components to ensure that they are harder to compromise or impair, (ii) \textit{Diversity} -- defined as the implementation of a feature with diverse types of components to restrict the threat impact from proliferating further into the system, and (iii) \textit{Principle of least privilege} -- defined as limiting the privileges of any entity, that is just enough to perform its functions and prevents the effect of threat from propagating beyond the affected component.
\begin{figure}
\centering
\includegraphics[width=0.7\linewidth]{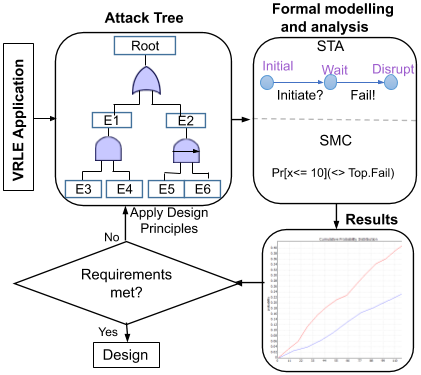}
\caption{\footnotesize{Proposed framework for security, privacy analysis in social VRLE.}}
\label{framework}
\vspace{-5mm}
\end{figure}
\section{Proposed framework}
\label{Sec:Problem}
In this section, we present details of our proposed framework that uses preliminary results from the previous section (Section III) for security and privacy analysis (i.e., to identify the most vulnerable attacks) of social VRLE applications. An overview of the steps followed in our framework is shown in Figure~\ref{framework}. Firstly, we outline the threat scenarios in a social VRLE application~\cite{vSocial} using traditional approaches. Secondly, we use an attack tree formalism for the modeling procedures. Following this, each attack tree is translated into an equivalent STA to form an NSTA, which is input into the UPPAAL SMC tool. Lastly, we use the quantitative assessment from the tool to determine if the probability of disruption is higher than a set threshold (user specified requirements). Based on this determination, we subsequently prescribe the design principles such as: \textit{hardening}, \textit{diversity} and \textit{principle of least privilege} that can be adopted in VRLE deployments. Overall, our framework steps help us to investigate potential security, privacy attack scenarios and recommend the design alternatives based on design principles for securing an edge computing based VRLE application.
\begin{figure*}[ht]
\small
\centering
\includegraphics[width=0.75\textwidth]{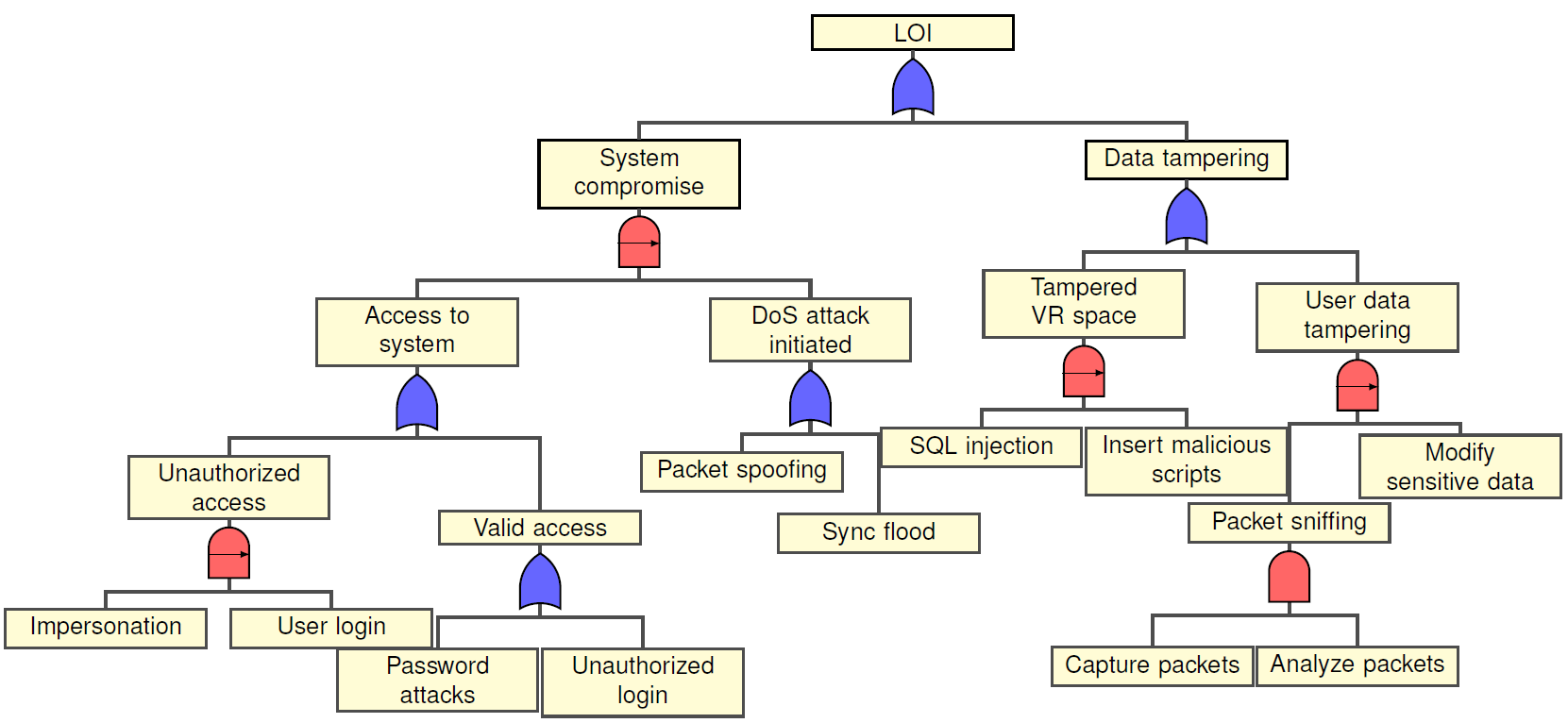}
\caption{\footnotesize{Formalized security attack tree with threat scenarios disrupting LoI.}}
\label{fig:AT1}
\end{figure*}

\vspace{-2.3mm}
\subsection{Formalization of security and privacy attack trees}
\label{attack_trees}
\noindent Attack trees are hierarchical models that show how an attacker goal (root node) can be refined into smaller sub-goals (child/intermediate nodes) via gates until no further refinement is possible such that the basic attack steps (BAS) are reached. BAS represents the \emph{leaf nodes} of an attack tree~\cite{PTA}. The \emph{leaf nodes} and the \emph{gates} connected in the attack trees are termed as \textit{attack tree elements}. To explore dependencies in attack surfaces, attack trees enable sharing of subtrees. Hence, attack trees are often considered as directed acyclic graphs, rather than trees~\cite{AFT}. 
\vspace{-2mm}
\begin{definition}[ \hspace{-2mm}Attack trees]
\label{AT_Formaldefinition}
An attack tree $A$ is defined as a tuple \{$N$,$Child$, $Top\_event$, $l$\} $\cup$ \{AT\_elements\}~where, 
$N$ is a finite set of nodes in the attack tree; 
Child: N $\rightarrow$ N* maps each set of nodes to its child nodes; 
Top\_event is an unique goal node of the attacker where 
Top\_event $\in$ N; 
l: is a set of labels for each node $n\in N$;
and AT\_elements: is a set of elements in an attack tree A. 
\end{definition} 
\noindent \textbf{Attack tree elements:} Attack tree elements aid in generating an attack tree and are defined as a set of $\{G \cup L\}$ where, $G$ represents gates; $L$ represents leaf nodes. Following are the descriptions of each of the AT elements.\\
\noindent \textbf{Attack tree gates:} Given an attack tree $A$, we formally define the attack tree gates $G = \{OR, AND, SAND\}$.\footnote{We limit our modeling to these three gates, however attack trees can adopt any other gates from the static/dynamic fault trees.}
An AND gate is disrupted when all its child nodes are disrupted, whereas an OR gate is disrupted if either of its child nodes are disrupted. Similarly, SAND gate is disrupted when all its child nodes are disrupted from left to right using the condition that the success of a previous step determines the success of the upcoming child node. The output nodes of the gates using these gates $G$ in an attack tree $A$ are defined as \emph{Intermediate nodes} $(I)$, which will be located at a level that is greater than the leaf nodes.
\newline
\noindent \textbf{Attack tree leaves:} An attack tree \emph{leaf node} is the terminal node with no other child node(s). It can be associated with \emph{basic attack steps (BAS)}, which collectively represent all the individual atomic steps within a composite attack scenario. To elucidate, for an attacker to perform intrusion, the prospective BAS can include: (i) identity spoofing, and (ii) unauthorized access to the system depending on the attacker profile. Thus, every BAS appears as an implicit leaf node of the attack tree. We assume the attack duration to have an exponential rate and model the equation as : $P(t) = 1-e^{\lambda t}$ where, $\lambda$ is the rate of exponential distribution. We use this exponential distribution because of its tractability and ease of handling, and also because it is defined by a single parameter.
\begin{figure*}
\small
\centering
\includegraphics[width=0.75\textwidth]{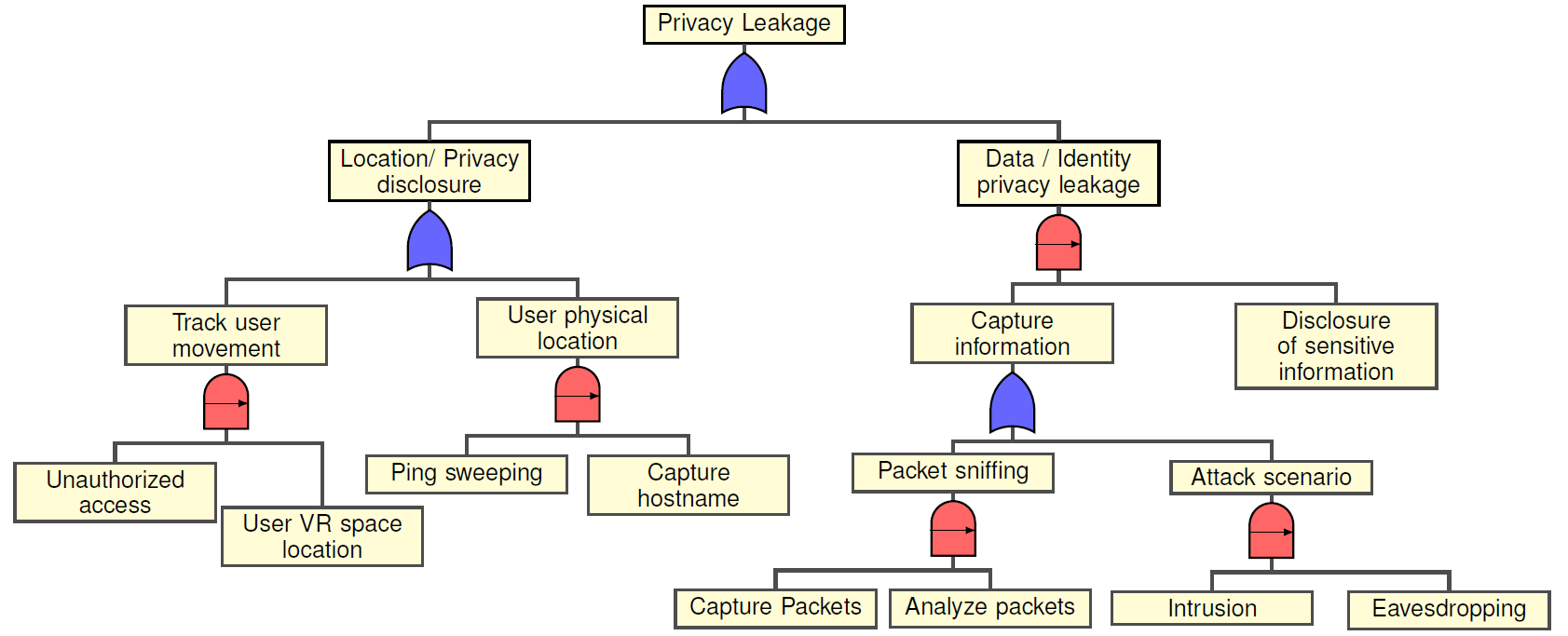}
\caption{\footnotesize{Formalized privacy attack tree with threat scenarios disrupting privacy leakage.}}
\label{fig:AT2}
\vspace{-6mm}
\end{figure*}
\\
\noindent \textbf{Security and privacy attack trees:} Based on the results discussed in Section~\ref{Sec:Preliminaries} and experimental evidence from our prior work \cite{CCNC}, we model threat scenarios in the form of a security attack tree (that lists potential VRLE security threats) and privacy attack tree (that lists potential VRLE privacy threats) as shown in Figures~\ref{fig:AT1} and~\ref{fig:AT2}, respectively. The descriptions of the leaf nodes are listed in Table~\ref{Table: SP-AT-table}. Exploring the security aspect in CIA triad of \{Confidentiality, Integrity, Availability\} may result into an enormous number of leaf nodes in the attack tree. Consequently, in this work we only focus on the Loss of Integrity (LoI) and privacy leakage to address the respective security and privacy threat scenarios that can disrupt the user experience in a social VRLE. Creation of new security and privacy trees for issues related to LoC and LoA can be performed similar to the approach presented for LoI; such details are beyond the scope of this paper.

Moreover, the listed attack trees are useful when they are concerned with user privacy and safety in a VRLE system. To elucidate, critical VRLE applications such as flight simulations, military training exercises and vSocial (developed for children with ASD) are sensitive to information disclosure attacks that can cause significant disruption for the stakeholder participants. If an attacker compromises such sensitive information, it can be used to harm the participants in the form of e.g., a chaperone attack (to make a user run into walls) and other physical safety attacks detailed in~\cite{new-haven}.
\vspace{-2mm}
\subsection{Translation of attack trees into stochastic timed automata}
\label{Sec:ATintoSTA}
In this section, we generate STA from the corresponding security and privacy attack trees shown in Figures~\ref{fig:AT1} and~\ref{fig:AT2}. In our translational approach: (i) each of the leaf nodes in these attack trees is converted into an individual STA. The intermediate events, which are basically the output of the logic gates used at different levels are converted imperatively into STA; (ii) the generated STAs are composed in parallel by including the root node; (iii) the obtained NSTA is then used for statistical model checking in order to verify the security and privacy properties formalized as SMC queries.
\begin{table*}[]
\small
\centering
\caption{\footnotesize{Descriptions of leaf nodes in security and privacy attack trees.}}
\resizebox{\textwidth}{!}{%
\begin{tabular}{|l|l|l|l|}
\hline
\multicolumn{2}{|c|}{\textbf{Security Attack Tree}} & \multicolumn{2}{c|}{\textbf{Privacy Attack Tree}} \\ \hline
\multicolumn{1}{|c|}{\textbf{Leaf Node Components}} & \multicolumn{1}{c|}{\textbf{Description of Leaf Nodes}} & \multicolumn{1}{c|}{\textbf{Leaf Node Components}} & \multicolumn{1}{c|}{\textbf{Description of Leaf Nodes}} \\ \hline
Impersonation & Attacker successfully assumes the identity of a valid user & Unauthorized Access & Attacker gains access to VR space \\ \hline
Packet Spoofing & Spoofing packets from a fake IP address to impersonate & User VR space location & \begin{tabular}[c]{@{}l@{}}Attacker determines the user location in\\ VR space\end{tabular} \\ \hline
Sync Flood & \begin{tabular}[c]{@{}l@{}}Sends sync request to a target and direct server resources \\ away from legitimate traffic\end{tabular} & Ping sweeping & \begin{tabular}[c]{@{}l@{}}Attacker sends pings to a range of IP \\ addresses and identify active hosts\end{tabular} \\ \hline
SQL Injection & \begin{tabular}[c]{@{}l@{}} Attacker injects malicious commands in user i/p query\\ using GET and POST\end{tabular} & Capture packets & \begin{tabular}[c]{@{}l@{}}Attacker uses packet sniffer to capture \\ packet information\end{tabular} \\ \hline
Insert Malicious Scripts & Attacker successfully adds malicious scripts in VR & Analyze packets & To identify erroneous packets to tamper \\ \hline
Capture Packets & \begin{tabular}[c]{@{}l@{}}The attacker uses a packet sniffer to capture packet \\ information\end{tabular} & Intrusion & \begin{tabular}[c]{@{}l@{}}Attacker performs an unauthorized activity\\ on VR space\end{tabular} \\ \hline
Analyze Packets & Attacker identifies erroneous packets to tamper & Eavesdropping & Attacker listens to conversations in VR space \\ \hline
Modify Sensitive Data & To modify any sensitive information by eavesdropping & \begin{tabular}[c]{@{}l@{}}Disclosure of sensitive\\ information\end{tabular} & Attacker maliciously releases any captured sensitive data \\ \hline
User Login & User login into VRLE & \multirow{3}{*}{Capture hostname} & \multirow{3}{*}{\begin{tabular}[c]{@{}l@{}}With IP address obtained, attacker can capture \\ the hostname in the VRLE application\end{tabular}} \\ \cline{1-2}
Unauthorized Login & Attacker gains access into VRLE by unauthorized means &  &  \\ \cline{1-2}
Password Attacks & Attacker recovers password of a valid-user &  &  \\ \hline
\end{tabular}}
\label{Table: SP-AT-table}
\end{table*}

\begin{figure}
\small
\centering
\includegraphics[width=0.24\textwidth]{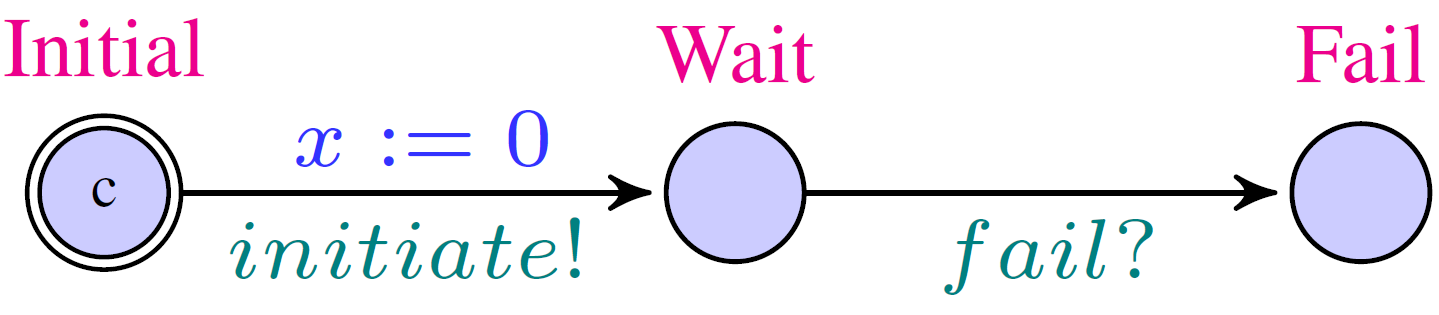}
\captionsetup{justification=centering}
\caption{\footnotesize{STA for root node.}}
\label{fig:TopEvent}
\vspace{-5mm}
\end{figure}

To demonstrate the translation of an attack tree into an STA, we consider the security attack tree as shown in Figure~\ref{fig:AT1}. As part of the translation, each of the security AT element (leaf and gates) input signals are connected to the output signal of child nodes. The generated network of STA communicates using signals. \textit{initiate} - indicates activation signal of attack tree element. This signal is sent initially from the root node to its children. \textit{fail} - indicates disruption of that attack tree element. This signal is sent to the parent node from its child node to indicate an STA disruption. The scope of the above signals can also be extended by special symbols such as: i)`?' (e.g., initiate?) means that the event will wait for the reception of the intended signal, ii) `!'(e.g., initiate!) implies output signal broadcasts to other STA in the attack tree.

\begin{figure}
\tiny
\input{STA_Tikz/OR1.tex}
\vspace{-2mm}
\captionsetup{justification=centering}
\caption{\footnotesize{STA for OR gate and root node of security attack tree.}}
\label{fig:Node_A_OR}
\vspace{-6mm}  
\end{figure}
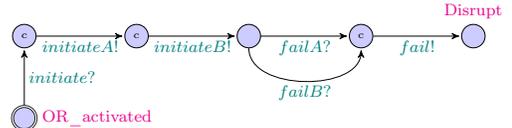 

\noindent \textit{\textbf{Illustrative example:}} 
For instance, we show the conversion of LoI i.e., root node ($Top\_event$) into STA as shown in Figure~\ref{fig:TopEvent}. The converted STA of the LoI is equipped with \emph{$initiate!$} and \emph{$fail?$} signals. The root node is the OR gate output for the two child nodes: (A) ``System Compromise'' and (B) ``Data Tampering''. Top OR gate sends an initiate signal and activates its child nodes ``System compromise'' and ``Data Tampering'' as shown in Figure~\ref{fig:Node_A_OR} by broadcasting $initiateA!$ and $initiateB!$ signals. After initialization, if either of the nodes (A) OR (B) are disrupted, then a $fail!$ signal is sent to the $Top\_event$, which forces a transition to $Disrupt$ state, representing LoI in the system. The clock $x$ is a UPPAAL global variable to keep track of the time progression as mentioned in Section~\ref{Sec:Preliminaries}. Similarly, STAs for the AND gate, SAND gates and the leaf nodes are also developed. Moreover, STAs for leaf nodes such as $impersonation$ in the security attack tree are instantiated with $\lambda$ (rate of exponential) values. For the given $\lambda$ values to the leaf nodes, the probability of occurrence is calculated. This value then propagates upward in the tree to calculate the probability of LoI. As mentioned earlier, the developed STAs are composed using the parallel composition~\cite{NSTA} technique to form an NSTA, which is then used for SMC by the UPPAAL tool~\cite{UPPAAL SMC Tutorial}. 


\section{Quantitative results}
\label{Sec:experiment}
\vspace{-1mm}
In this section, we present the results obtained using our proposed framework. As mentioned in Section~\ref{Sec:Problem}, the threat scenarios we consider are: \textit{LoI} and \textit{privacy leakage} for security and privacy attack trees (AT), respectively. In the following analysis, we assume that our design requirement is to keep the probability of LoI and privacy leakage below the threshold of 0.25. For evaluation purposes, we use arbitrary values of $\lambda$ as parametric input to the leaf nodes as shown in Table~\ref{table:metrics} obtained from~\cite{QUIRC},~\cite{SQL_Injection}.
\begin{table}
\centering
\caption{\footnotesize{$\lambda$ values for leaf nodes of security \& privacy ATs.}}
\small
\resizebox{0.5\textwidth}{!}{%
\begin{tabular}{|c|c|c|c|}
\hline
\multicolumn{2}{|c|}{\textbf{Security AT}} & \multicolumn{2}{c|}{\textbf{Privacy AT}} \\ \hline
\textbf{Security threats} & \textbf{$\lambda$} & \textbf{Privacy threats} & \textbf{$\lambda$} \\ \hline
Impersonation & 0.006892 & Unauthorized access & 0.006478 \\ \hline
User login & 0.0089 & User VR space location & 0.0094 \\ \hline
Password attacks & 0.008687 & Capture hostname & 0.004162 \\ \hline
Unauthorized login & 0.008687 & Ping sweeping & 0.002162 \\ \hline
Packet spoofing & 0.0068 & Capture packets & 0.00098 \\ \hline
SYNC flood & 0.0068 & Analyze packets& 0.0048 \\ \hline
SQL injection & 0.00231788 & Disclosure of sensitive info & 0.0009298 \\ \hline
Insert malicious scripts & 0.008 & Intrusion & 0.006628 \\ \hline
Capture packets & 0.000 98 & Eavesdropping & 0.08 \\ \hline
Analyze packets & 0.0048 & -- & -- \\ \hline
Modify sensitive data & 0.002642 & -- & -- \\ \hline
\end{tabular}%
}
\label{table:metrics}
\vspace{-4.5mm}
\end{table}
Note, after providing $\lambda$ values as parameters to the leaf nodes, we utilize the SMC queries as explained in Section~\ref{STA-ex} to find the respective probabilities of LoI and privacy leakage. Any other user specified threshold values can also be used in our framework. This is due to the fact that the model checking approach takes the user specified values at the beginning of an experiment. For our experiment purposes, we consider LoI (security attack tree) and privacy leakage (privacy attack tree) as goal nodes. In the following set of experiments, we present the obtained probability of the goal nodes with respect to the time window used by the attacker.
\vspace{-1mm}
\subsection{Vulnerability analysis in the security AT}
\vspace{-1.5mm}
\label{Analysis-securityAT}

\noindent We assign the values of $\lambda$ shown in Table~\ref{table:metrics}. However, when assigning a $\lambda$ value to a leaf node in the attack tree, we consider a very small positive constant  ($K$) $\approx$ 0.002 for the remaining leaf nodes. This is because, in real time systems, multiple attack scenarios can happen.
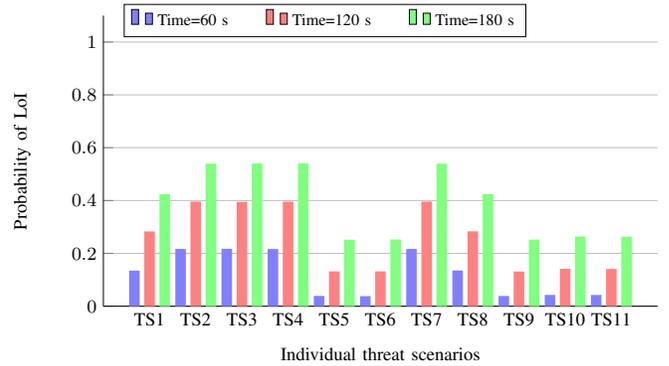
\begin{figure}[t]
\footnotesize
\input{Parallel_Figures/Security_Individual.tex}
\vspace{-3mm}
\captionsetup{justification=centering}
\caption{\footnotesize{TS of security AT where - $TS2$, $TS3$, $TS4$, $TS7$ are the most vulnerable nodes.}}
\label{fig:Sec_Pr_Leaf}
\vspace{-4mm}
\end{figure}
To identify a vulnerability in a security attack tree, we analyze: (i) individual leaf nodes, and (ii) combinations of leaf nodes, to determine their effect on the probability of LoI occurrence.

\noindent \textit{\textbf{i) Individual leaf node analysis:}}
\noindent 
In Figure~\ref{fig:Sec_Pr_Leaf}, we show the probability of LoI over multiple time windows for each leaf node in the security attack tree. We perform a thorough analysis of leaf nodes in the security attack tree for threat scenarios across different time intervals i.e., \textit{t = \{0, 60, 120, 180\}}. For the individual leaf node analysis, the considered threat scenarios (TS) shown in Figure~\ref{fig:Sec_Pr_Leaf} are termed as: $TS1$ -- insert malicious scripts, $TS2$ -- packet spoofing, $TS3$ -- unauthorized login,  $TS4$ -- password attacks, $TS5$ -- modify data, $TS6$ -- analyze packets, $TS7$ -- Sync flood, $TS8$ -- SQL injection, $TS9$ -- capture packets, $TS10$ -- impersonation, $TS11$ -- user login. As shown in Figure~\ref{fig:Sec_Pr_Leaf}, the leaf nodes $TS3$ and $TS4$ (for unauthorized access) as well as $TS2$ and $TS7$ (for DoS attack) are the most vulnerable in the security attack tree with the probability of 0.53.
\begin{figure}[t]
\footnotesize
\input{Parallel_Figures/Security_combinations.tex}
\vspace{-2.8mm}
  \captionsetup{justification=centering}
  \caption{\footnotesize{TS of security AT where -- $TS6\superscript*$, $TS7\superscript*$ are the most vulnerable combination.}}
    \label{Fig:Security_Combinations}
\vspace{-3mm}
\end{figure}
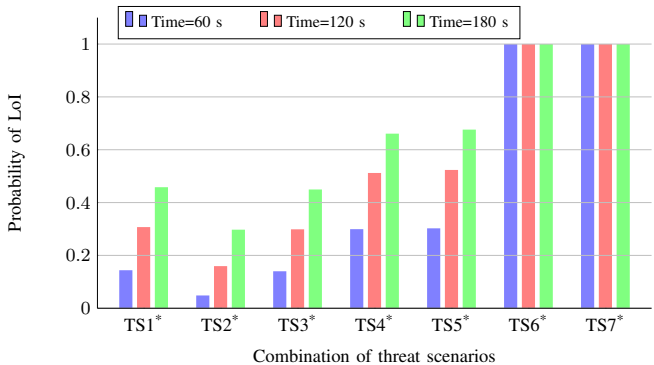

\noindent \textbf{\textit{ii) Analysis using combination of leaf nodes:}} Herein, we consider combinations of leaf nodes to identify their impact on LoI. For these experiments, we explore two scenarios: In the first scenario, we consider combinations of leaf nodes that belong to the same sub-tree, and in the second scenario, we consider leaf nodes from different sub-trees. The considered combination of threat scenarios are enlisted as: $TS1\superscript* $ -- \{impersonation, SQL injection\}, $TS2\superscript*$ -- \{impersonation, modify data\}, $TS3\superscript*$ -- \{SQL injection, capture packets\}, $TS4\superscript*$ -- \{pwd attacks, SQL injection\}, $TS5\superscript*$ -- \{impersonation, packet spoofing\}, $TS6\superscript*$ -- \{packet spoofing, unauthorized login\}, $TS7\superscript*$ -- \{unauthorized login, Sync flood\}. As shown in Figure~\ref{Fig:Security_Combinations}, $TS6\superscript*$ and $TS7\superscript*$ are the most vulnerable combination of threat scenarios with a probability of 1 for an LoI event. As part of further analysis in Section~\ref{Sec:Design-principles}, we discuss about the potential candidates for design principles to apply on these leaf nodes such that the VRLE application resilience against security threats is enhanced. 
\vspace{-2mm}
\subsection{Vulnerability analysis in the privacy AT}
\label{Analysis-privacyAT}
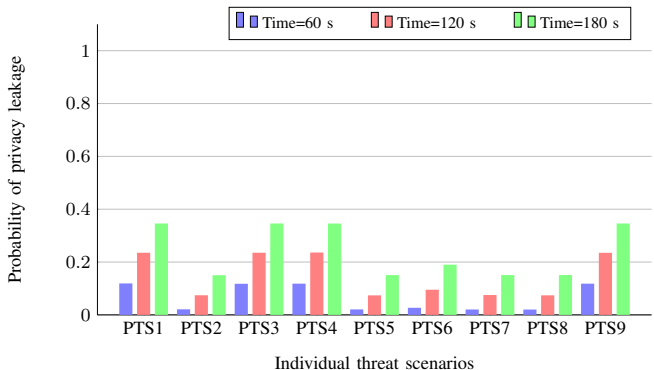
\begin{figure}[t]
\footnotesize
\input{Parallel_Figures/Privacy_Individual.tex}
\vspace{-3mm}
\captionsetup{justification=centering}
\caption{\footnotesize{TS of privacy AT where - $PTS1$, $PTS3$, $PTS4$, $PTS9$ are the most vulnerable nodes.}}
\label{Privacy_Individual}
\vspace{-4mm}
\end{figure}
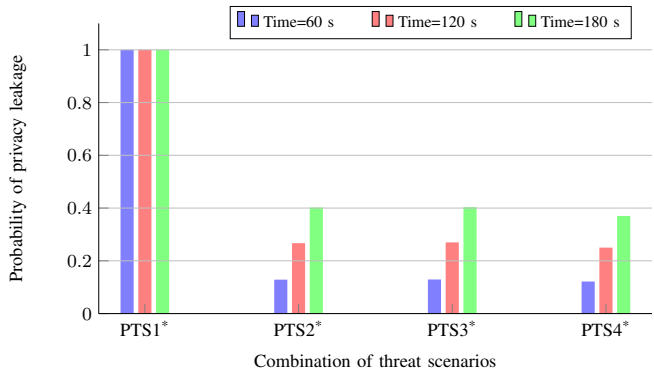
\begin{figure}
\footnotesize
\input{Parallel_Figures/Privacy_combinations.tex}
\captionsetup{justification=centering}
\caption{\footnotesize{TS of privacy AT where -- $PTS1\superscript*$ is the most vulnerable combination.}}
\label{Privacy_Combinations}
\vspace{-5mm}
\end{figure}
\begin{table*}
\centering
\caption{\footnotesize{Most vulnerable components considering the individual \& combination of leaf nodes.}}
\resizebox{0.9\textwidth}{!}{%
\begin{tabular}{|c|c|c|c|c|}
\hline
\multirow{2}{*}{\textbf{\begin{tabular}[c]{@{}c@{}}Level in \\ attack\\ trees\end{tabular}}} & \multicolumn{2}{c|}{\textbf{Analysis on security AT}} & \multicolumn{2}{c|}{\textbf{Analysis on privacy AT}} \\ \cline{2-5} 
 & \textbf{\begin{tabular}[c]{@{}c@{}}Different\\ Scenarios\end{tabular}} & \textbf{\begin{tabular}[c]{@{}c@{}}Identified vulnerable\\ components in security \\ AT\end{tabular}} & \textbf{\begin{tabular}[c]{@{}c@{}}Different\\ Scenarios\end{tabular}} & \textbf{\begin{tabular}[c]{@{}c@{}}Identified vulnerable\\ components in privacy \\ AT\end{tabular}} \\ \hline
\multirow{4}{*}{\textbf{\begin{tabular}[c]{@{}c@{}}Individual\\ leaf nodes\end{tabular}}} & \multirow{4}{*}{\begin{tabular}[c]{@{}c@{}}Leaf nodes where\\ probability of\\ disruption in LoI at\\ (t\leq180) = 0.53\end{tabular}} & Unauthorized login & \multirow{4}{*}{\begin{tabular}[c]{@{}c@{}}Leaf nodes where\\ probability of\\ disruption in privacy leakage \\ at (t\leq180) = 0.34\end{tabular}} & Unauthorized access \\ \cline{3-3} \cline{5-5} 
 &  & Packet spoofing &  & User VR space location \\ \cline{3-3} \cline{5-5} 
 &  & Sync flood &  & Ping sweeping \\ \cline{3-3} \cline{5-5} 
 &  & Password attacks &  & Capture hostname \\ \hline
\textbf{\begin{tabular}[c]{@{}c@{}}Combinat\\ -ion of leaf\\ nodes\end{tabular}} & \begin{tabular}[c]{@{}c@{}}Leaf nodes where\\ probability of\\ disruption in LoI \\ at (t\leq180) = 1\end{tabular} & \begin{tabular}[c]{@{}c@{}}\{Unauthorized login, \\ Packet spoofing\}, \\ \{Unauthorized login,\\ Sync flood\}\end{tabular} & \begin{tabular}[c]{@{}c@{}}Leaf nodes where\\ probability of\\ disruption in privacy leakage at\\ (t\leq180) = 1\end{tabular} & \begin{tabular}[c]{@{}c@{}}\{Unauthorized access,\\ user VR space location\}\end{tabular} \\ \hline
\end{tabular}}
\captionsetup{justification=centering}
\vspace{-7mm}
\label{Table: Results}
\end{table*}
\noindent We analyze the privacy attack tree similarly for: (i) individual leaf nodes, and (ii) combinations of leaf nodes. For the considered individual leaf node analysis in the privacy attack tree, the threat scenarios are termed as: $PTS1$ -- unauthorized access, $PTS2$ -- capture packets, $PTS3$ -- user VR space location, $PTS4$ -- ping sweeping, $PTS5$ -- analyze packets, $PTS6$ -- disclosure of sensitive information, $PTS7$ -- intrusion, $PTS8$ -- eavesdropping, $PTS9$ -- capture hostname. As shown in Figure~\ref{Privacy_Individual}, the most vulnerable leaf nodes are: $PTS1$, $PTS3$, $PTS4$, $PTS9$ with the highest probability of privacy leakage of 0.34.

For the analysis of combination of leaf nodes, we refer to the combination of threat scenarios as: $PTS1$\superscript* -- \{unauthorized access, user VR space location\}, $PTS2$\superscript*-- \{capture packets, disclosure of sensitive information\}, $PTS3$\superscript* -- \{unauthorized access, disclosure of sensitive information\}, $PTS4$\superscript* -- \{capture packets, analyze packets\}. As shown in Figure~\ref{Privacy_Combinations}, $PTS1$\superscript* is the most vulnerable combination of threats for privacy leakage with a probability of 1. In summary, we can conclude that the above numerical analysis shown in Table~\ref{Table: Results} on both security and privacy attack trees can help in identifying the LoI and privacy leakage concerns that need to be addressed in the social VRLE design.

\section{Recommended Design Principles}
\label{Sec:Design-principles}
\noindent 
In this section, we are examining the effect of applying various design principles to the most vulnerable components identified in the Sections~\ref{Analysis-securityAT}, and \ref{Analysis-privacyAT}. Existing works such as NIST SP800-160 document~\cite{NIST},~\cite{Designprinciples} suggest that the services for safeguarding security and privacy are critical for successful operation of current devices and sensors connected to physical networks as part of IoT systems. As mentioned in Section~\ref{definition-design-principles}, these design principles are essential to construct a trustworthy edge computing based system architecture. The goal is to apply a combination of design principles at different levels of abstraction to help in developing effective mitigation strategies. We adopt a selection of design principles such as  \textit{hardening}, \textit{diversity} and \textit{principle of least privilege} among the list of principles available in NIST document~\cite{NIST}, and~\cite{Designprinciples}. In the following, we demonstrate their effectiveness by showing that there is a reduction in the probability of disruption terms after adopting them in our VRLE system design. 
\begin{figure}
\footnotesize
\input{Four_Diagrams/Security_Princ.tex}
\vspace{-3mm}
\captionsetup{justification=centering}
\caption{\footnotesize{Prob. in LoI reduced by 15.85\% in security AT due to application of design principles.}}
\label{Fig:Security_Hardening}
\vspace{-4mm}
\end{figure}
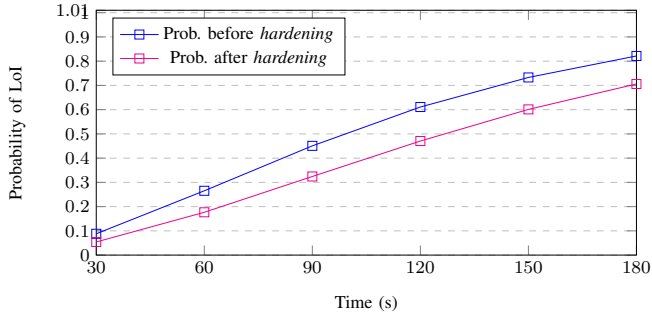
\newline
\noindent \textit{\textbf{Implementation of design principles on security attack tree:}}
In this section, we apply design principles on one of the identified vulnerable nodes of the security attack tree as shown in Section \ref{Analysis-securityAT}. For instance, we incorporate \textit{hardening} design principle on the password attacks, to study its effects on the security metric LoI as shown in Figure~\ref{Fig:Security_Hardening}. As part of the \textit{hardening} principle, we added new nodes such as a firewall and a security protocol in the security attack tree. Our results show that the probability of disruption of LoI is reduced from 0.82 to 0.69 (15.85\%), with the given attacker profile. The decrease in the disruption of LoI is due to the rise in additional resources that are required by the attacker to compromise such a VRLE application system which is incorporating the \textit{hardening} principle. Similarly, we apply the \textit{principle of least privilege} on the security attack tree, which under-provisions privileges intentionally. This in turn reduced the probability of disruption of LoI from 0.82 to 0.79 (3.66\%). \\


\begin{figure}
\footnotesize
\input{Four_Diagrams/Privacy_Princ.tex}
\vspace{-3mm}
\captionsetup{justification=centering}
\caption{\footnotesize{Prob. of privacy leakage reduced by 68\% in privacy AT due to application of design principles.}}
\label{Diversity}
\vspace{-6mm}
\end{figure}
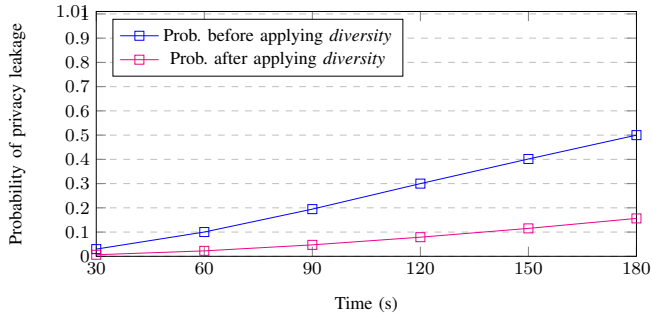
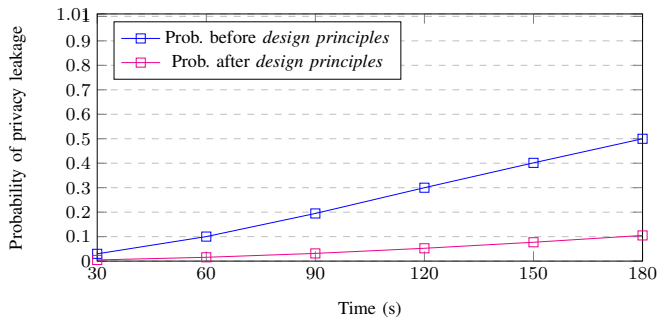
\begin{figure}
\footnotesize
\input{Four_Diagrams/Privacy_Combined.tex}
\vspace{-3.5mm}
\captionsetup{justification=centering}
\caption{\footnotesize{Prob. of privacy leakage reduced by 80\% in privacy AT due to application of design principles for a combination of privacy AT nodes.}}
\label{Privacy_All_Defence}
\vspace{-4mm}
\end{figure}

\noindent \textit{\textbf{Implementation of design principles on privacy attack tree:}}
Using the similar approach mentioned in design principles on the security attack tree, we apply \textit{diversity} design principle on one of the identified vulnerable nodes (unauthorized access) in the privacy attack tree. After adding multi-factor authentication procedures as part of the \textit{diversity} principle, the probability of disruption on privacy leakage is reduced significantly from 0.5 to 0.16 (68\%) as shown in Figure~\ref{Diversity}. Similarly, we apply the \textit{principle of least privilege} by under-provisioning privileges on the privacy attack tree where the probability of disruption of privacy leakage is slightly reduced from 0.5 to 0.48 (4\%). Thus, from the above implementation of individual design principles, we conclude that \textit{hardening} and \textit{diversity} are more effective in reducing the disruption of LoI and privacy leakage, respectively.
Thus, our findings shows that some security principles are more effective than others. In addition, our results emphasize the benefits in implementing a combination of design principles in both security and privacy attack trees to overall improve the attack mitigation efforts. 
\begin{figure}
\footnotesize
\input{Four_Diagrams/Security_Combined.tex}
\vspace{-3.5mm}
\captionsetup{justification=centering}
\caption{\footnotesize{Prob. of LoI is reduced by 26\% in security AT due to application of design principles for a combination of security attack tree nodes.}}
\label{Security_All_Defence}
\vspace{-6mm}
\end{figure}
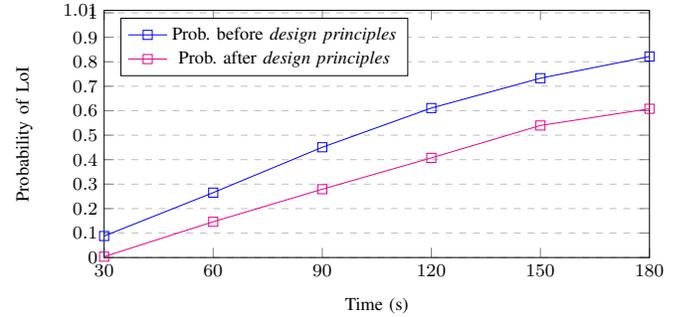
\newline
\indent To study the effect on disruption of the LoI and privacy leakage, we adopt a combination of design principles such as: (i) for the security attack tree: \{\textit{hardening}, \textit{principle of least privilege}\}, and (ii) for the privacy attack tree: \{\textit{diversity}, \textit{principle of least privilege}\}. We observe that there is a significant drop in the probability of disruption of LoI from 0.81 to 0.6 (26\%), and 0.5 to 0.1 (80\%) for privacy leakage as shown in Figures~\ref{Security_All_Defence} and~\ref{Privacy_All_Defence}, respectively.
\newline
\indent From the above numerical analysis, we can conclude that incorporating relevant combination of standardized design principles and their joint implementation have the potential to better mitigate the impact of sophisticated and well-orchestrated cyber attacks on edge computing assisted VRLE systems with IoT devices. In addition, our above results provide insights on how the adoption of the design principles can provide the necessary evidence to support a trustworthy level of security and privacy for the users in VRLE systems that are used for important societal applications such as: special education, surgical training, and flight simulators.

\section{Conclusion}
\label{Sec:conclusion}
Social Virtual Reality Learning Environments (VRLEs) are a new form of immersive VR applications, where security and privacy issues are under-explored. In this paper, we presented a novel framework that quantitatively assesses the security and privacy threat scenarios for a social VRLE application case study viz., vSocial. Specifically, we explored different threat scenarios that possibly cause LoI (e.g., unauthorized access) and privacy leakage (e.g., disclosure of sensitive user information) in a set of social VRLE application session scenarios. We utilized the attack tree formalism to model the security and privacy threats. Specifically, we developed relevant attack trees and converted them into stochastic timed automata and then performed statistical model checking using the UPPAAL SMC tool. Furthermore, we illustrated the effectiveness of our framework by analyzing different design principle candidates. We showed a `before' and `after' performance comparison to investigate the effect of applying these design principles on the probability of LoI and privacy leakage occurrence. The highlights from our experiments with realistic social VRLE application scenarios indicate that some security principles are more effective than others. However, combining them can result in a more effective mitigation mechanism. For instance, among the design principle candidates, (i) \textit{\{\textit{hardening}, \textit{principle of least privilege}\}} is the best design principle combination for enhancing security, and (ii) \textit{\{\textit{diversity}, \textit{principle of least privilege}\}} is the best design principle combination for enhancing privacy.

In future, we plan to explore the effect of fault and attacks as a combination using the attack-fault tree formalism~\cite{AFT} for VRLE applications. This will allow us to reason about the safety metrics and study the safety, security and privacy trade-offs. Since, different components in a typical social VRLE application go through different maintenance actions, we also plan to explore the impact of various maintenance strategies on the reliability metric of social VRLE applications using the fault maintenance tree formalism~\cite{SmartFT}.

\vspace{12pt}
\end{document}

%% file: STA_Tikz/Top_Event.tex
\begin{tikzpicture}[>=stealth',
    shorten > = 1pt,
node distance = 0.8cm and 1.6cm,
    el/.style = {inner sep=2pt, align=left, sloped}
                    ]

\node (q0) [state, accepting, label=above:{\color{magenta}\footnotesize{Initial}}, fill=blue!20, minimum size=0.3cm]  {};

\node (q1) [state,right=of q0, label=above:{\color{magenta}\footnotesize{Wait}}, fill=blue!20, {}, minimum size=0.3cm] {};
\node (q2) [state,right=of q1, label=above:{\color{magenta}\footnotesize{Fail}}, fill=blue!20, {}, minimum size=0.3cm] {};

\path[->]    (q0)  edge [right=10]  node[el,below]  {\color{teal}\footnotesize{$initiate!$ }} node[el,above]  {\color{blue!80}\footnotesize{$x:=0$}}        (q1)

   
    (q1)  edge [right=10]  node[el,below]  {\color{teal}\footnotesize{$fail?$}}         (q2);
\end{tikzpicture}

%% file: STA_Tikz/OR1.tex
\centering
\resizebox{0.75\linewidth}{!}{      
\begin{tikzpicture}[>=stealth',
    shorten > = 1pt,
node distance = 1.4cm and 1.5cm,
    el/.style = {inner sep=2pt, align=left, sloped}
                    ]
                                        \fontfamily{Aerial};

\node (q1) [state,right=of q0, label=above:{}, fill=blue!20, {}, minimum size=0.4cm] {\tiny{c}};
\node (q0) [state, accepting, below of =q1, label=right:{\color{magenta}\footnotesize{{OR\_activated}}}, fill=blue!20, minimum size=0.4cm]  {};

\node (q2) [state,right=of q1, fill=blue!20, {}, minimum size=0.4cm] {\tiny{c}};
\node (q3) [state,right=of q2, fill=blue!20, {}, minimum size=0.4cm] {};
\node (q4) [state,right=of q3, fill=blue!20, {}, minimum size=0.4cm] {\tiny{c}};
\node (q5) [state,right=of q4, fill=blue!20, label=above:{\color{magenta}\footnotesize{{Disrupt}}}, minimum size=0.4cm] {};


\path[->]    (q0)  edge [right=10]  node[el,right, rotate=270] {\color{teal}\footnotesize{$initiate?$}}         (q1)
    (q1)  edge [right=10]  node[el,below]  {\color{teal}\footnotesize{$initiateA!$}}         (q2)

(q2)  edge [right=10]  node[el,below]  {\color{teal}\footnotesize{$initiateB!$}}         (q3)
(q3)  edge [right=10]  node[el,below]  {\color{teal}\footnotesize{$failA?$}}         (q4)
(q3)  edge [bend right=90]  node[el,below]  {\color{teal}\footnotesize{$failB?$}}     (q4)
(q4)  edge [right=10]  node[el,below]  {\color{teal}\footnotesize{$fail!$}}         (q5);
\end{tikzpicture}} 

%% file: Parallel_Figures/Security_Individual.tex
\centering
\resizebox{\linewidth}{!}{      
\begin{tikzpicture}[scale=0.8]
  \centering
  \begin{axis}[
        ybar,
        bar width=0.16cm,
        width=10cm,height=6cm,
        ymajorgrids, tick align=inside,
        enlarge y limits={value=.1,upper},
        ymin=0, ymax=1,
        ytick={0, 0.2, 0.4, 0.6, 0.8, 1},
        axis x line*=bottom,
        axis y line*=left,
        legend style={
            at={(0.4, 1.04)},
            anchor=north,
            legend columns=-1,
            /tikz/every even column/.append style={column sep=0.5cm}
        },
        ylabel={\footnotesize{Probability of LoI}},
        xlabel={\footnotesize{Individual threat scenarios}},
        symbolic x coords= {TS1, TS2, TS3, TS4, TS5, TS6, TS7, TS8, TS9, TS10, TS11},
       xtick=data,
    ]
    \addplot [draw=none, fill=blue!50] coordinates {
       (TS1,0.134783) (TS2,0.216752) (TS3,0.217211) (TS4,0.216554) (TS5,0.038892) (TS6,0.0380003) (TS7,0.216752) (TS8,0.135194) (TS9,0.03871) (TS10,0.0427413) (TS11, 0.0426483) };
   \addplot [draw=none,fill=red!50] coordinates {
      (TS1,0.283075) (TS2,0.396105)  (TS3,0.39491) (TS4,0.395731) (TS5,0.131918) (TS6,0.131961) (TS7,0.396105) (TS8,0.283393) (TS9,0.13131) (TS10,0.141369) (TS11,0.141016) };
   \addplot [draw=none, fill=green!50] coordinates {
       (TS1,0.423916) (TS2,0.539493)(TS3,0.540115) (TS4,0.540439) (TS5,0.25168) (TS6,0.252497) (TS7,0.539493) (TS8,0.424269) (TS9,0.252143) (TS10,0.263724) (TS11,0.263071) };

    \legend{\scriptsize{Time=60 s}, \scriptsize{Time=120 s},\scriptsize{Time=180 s}}
\end{axis}
\end{tikzpicture}
}       

%% file: Parallel_Figures/Security_combinations.tex
\centering
\resizebox{\linewidth}{!}{      
\begin{tikzpicture}[scale=0.8]
  \centering
  \begin{axis}[
        ybar, axis on top,
        bar width=0.2cm,                width=10cm, height=6cm,
        ymajorgrids, tick align=inside,
        enlarge y limits={value=.1,upper},
        ymin=0, ymax=1,
        ytick={0, 0.2, 0.4, 0.6, 0.8, 1},
        axis x line*=bottom,
        axis y line*=left,
        y axis line style={opacity=1},
        tickwidth=0pt,
        legend style={
            at={(0.4,1.04)},
            anchor=north,
            legend columns=-1,
            /tikz/every even column/.append style={column sep=0.5cm}
        },
        ylabel=\footnotesize{{Probability of LoI}},
        xlabel=\footnotesize{{Combination of threat scenarios}},
        symbolic x coords={
        TS1\superscript*, TS2\superscript*, TS3\superscript*, TS4\superscript*, TS5\superscript*, TS6\superscript*, TS7\superscript*
           },
       xtick=data
    ]
    \addplot [draw=none, fill=blue!50] coordinates {
      (TS1\superscript*, 0.143867) (TS2\superscript*,0.0483566) (TS3\superscript*, 0.139935) (TS4\superscript*, 0.299295) (TS5\superscript*, 0.302461) (TS6\superscript*, 1) (TS7\superscript*, 1) };
   \addplot [draw=none,fill=red!50] coordinates {
      (TS1\superscript*, 0.307153) (TS2\superscript*,0.159412) (TS3\superscript*, 0.298732) (TS4\superscript*, 0.512037) (TS5\superscript*, 0.523459) (TS6\superscript*, 1) (TS7\superscript*, 1)  };
   \addplot [draw=none, fill=green!50] coordinates {
      (TS1\superscript*, 0.458105) (TS2\superscript*, 0.297356) (TS3\superscript*, 0.449516) (TS4\superscript*, 0.660812) (TS5\superscript*, 0.676113) (TS6\superscript*, 1) (TS7\superscript*, 1)       };

    \legend{\scriptsize{Time=60 s},\scriptsize{Time=120 s},\scriptsize{Time=180 s}}
  \end{axis}
  \end{tikzpicture}
}       

%% file: Parallel_Figures/Privacy_Individual.tex
\centering
\resizebox{\linewidth}{!}{      
\begin{tikzpicture}[scale=0.8]
  \centering
  \begin{axis}[
        ybar,
        width=10cm,height=6cm,
        bar width=0.2cm,
        ymajorgrids, tick align=inside,
        enlarge y limits={value=.1,upper},
        ymin=0, ymax=1,
        ytick={0, 0.2, 0.4, 0.6, 0.8, 1},
        axis x line*=bottom,
        axis y line*=left,
        y axis line style={opacity=1},
        tickwidth=0pt,
        enlarge x limits=true,
        legend style={
            at={(0.6, 1.06)},
            anchor=north,
            legend columns=-1,
            /tikz/every even column/.append style={column sep=0.5cm}
        },
        ylabel={\footnotesize{Probability of privacy leakage}},
        xlabel={\footnotesize{Individual threat scenarios}},
        symbolic x coords={ PTS1, PTS2, PTS3, PTS4, PTS5, PTS6, PTS7, PTS8, PTS9},
       xtick=data
       ]
    \addplot [draw=none, fill=blue!50] coordinates {
       (PTS1, 0.119059) (PTS2, 0.0210912)  (PTS3, 0.117733) (PTS4, 0.118155) (PTS5, 0.0206553) (PTS6, 0.0268139) (PTS7, 0.0204441) (PTS8, 0.020203)  (PTS9, 0.118174)};
   \addplot [draw=none,fill=red!50] coordinates {
      (PTS1, 0.235215) (PTS2, 0.0742386) (PTS3, 0.235223) (PTS4, 0.235941) (PTS5, 0.0738291) (PTS6, 0.0952278) (PTS7, 0.0750692) (PTS8, 0.073951) (PTS9, 0.234974) };
   \addplot [draw=none, fill=green!50] coordinates {
       (PTS1, 0.345908) (PTS2, 0.150028) (PTS3, 0.345914) (PTS4, 0.345671)  (PTS5, 0.15063) (PTS6, 0.190143) (PTS7, 0.150633) (PTS8, 0.150764) (PTS9, 0.345805) };
    \legend{\scriptsize{Time=60 s},\scriptsize{Time=120 s},\scriptsize{Time=180 s}}
  \end{axis}
  \end{tikzpicture}
}       

%% file: Parallel_Figures/Privacy_combinations.tex
\centering
\resizebox{\linewidth}{!}{      
\begin{tikzpicture}[scale=0.75]
  \centering
  \begin{axis}[
        ybar, axis on top,
        width=10cm,height=6cm,
        bar width=0.2cm,
        ymajorgrids, tick align=inside,
        enlarge y limits={value=.1,upper},
        ymin=0, ymax=1,
        ytick={0, 0.2, 0.4, 0.6, 0.8, 1},
        axis x line*=bottom,
        axis y line*=left,
        y axis line style={opacity=1},
        enlarge x limits=true,
        legend style={
            at={(0.6, 1.06)},
            anchor=north,
            legend columns=-1,
            /tikz/every even column/.append style={column sep=0.5cm}
        },
        ylabel=\footnotesize{{Probability of privacy leakage}},
        xlabel=\footnotesize{{Combination of threat scenarios}},
        symbolic x coords={
           PTS1\superscript*, PTS2\superscript*, PTS3\superscript*, PTS4\superscript*},
       xtick=data
       ]
    \addplot [draw=none, fill=blue!50] coordinates {
      (PTS1\superscript*,1) (PTS2\superscript*, 0.128578) (PTS3\superscript*, 0.129261)  (PTS4\superscript*, 0.121516)
      };
   \addplot [draw=none,fill=red!50] coordinates {
      (PTS1\superscript*, 1) (PTS2\superscript*, 0.266997) (PTS3\superscript*, 0.269792 )  (PTS4\superscript*, 0.250057)
      };
   \addplot [draw=none, fill=green!50] coordinates {
      (PTS1\superscript*, 1) (PTS2\superscript*, 0.402067) (PTS3\superscript*, 0.403694)  (PTS4\superscript*, 0.370003)
      };

    \legend{\scriptsize{Time=60 s},\scriptsize{Time=120 s},\scriptsize{Time=180 s}}
  \end{axis}
  \end{tikzpicture}
}       

%% file: Four_Diagrams/Security_Princ.tex
\centering
\resizebox{\linewidth}{!}{      
\begin{tikzpicture}
\begin{axis}[
    xlabel={Time (s)},
    ylabel={Probability of LoI},
      width=10cm, height=5.4cm,
    xmin=30, xmax=180,
    ymin=0, ymax=1.01,
    xtick={30, 60, 90 ,120, 150 ,180},
    ytick={0,0.1,0.2,0.3,0.4,0.5,0.6,0.7,0.8,0.9,1.0,1.01},
    legend pos=north west,
    ymajorgrids=true,
    grid style=dashed,
]
 
\addplot[
    color=blue,
    mark=square,
    ]
    coordinates {
    (30,0.087979) (60,0.265287) (90,0.450649) (120, 0.610879) (150, 0.732862) (180, 0.821445)
};
\addplot[
    color=magenta,
    mark=square,
    ]
    coordinates {
    (30,0.0541373) (60,0.176598) (90, 0.324313) (120, 0.470547) (150, 0.601087) (180, 0.705879)
};
    
    \legend{Prob. before \textit{hardening}, Prob. after \textit{hardening}}
\end{axis}
\end{tikzpicture}
}       

%% file: Four_Diagrams/Privacy_Princ.tex
\centering
\resizebox{\linewidth}{!}
{ 
\begin{tikzpicture}
\begin{axis}[
    xlabel={Time (s)},
    ylabel={Probability of privacy leakage},
      width=10cm, height=5.4cm,
    xmin=30, xmax=180,
    ymin=0, ymax=1.01,
    xtick={30, 60, 90 ,120, 150 ,180},
    ytick={0,0.1,0.2,0.3,0.4,0.5,0.6,0.7,0.8,0.9,1.0,1.01},
    legend pos=north west,
    ymajorgrids=true,
    grid style=dashed,
]
 
\addplot[
    color=blue,
    mark=square,
    ]
    coordinates {
    (30,0.0297068) (60,0.100165) (90,0.194637) (120, 0.299908) (150, 0.401481) (180, 0.499895)
};
\addplot[
    color=magenta,
    mark=square,
    ]
    coordinates {
    (30,0.00642001) (60, 0.0224657) (90, 0.0472021) (120, 0.0788827) (150, 0.115184) (180, 0.15636)
};
    
    \legend{Prob. before applying \textit{diversity}, Prob. after applying \textit{diversity}}
\end{axis}
\end{tikzpicture}
}

%% file: Four_Diagrams/Security_Combined.tex
\centering
\resizebox{\linewidth}{!}{      
\begin{tikzpicture}
\begin{axis}[
    xlabel={Time (s)},
    ylabel={Probability of LoI},
      width=10cm, height=5.4cm,
    xmin=30, xmax=180,
    ymin=0, ymax=1.01,
    xtick={30, 60, 90 ,120, 150 ,180},
    ytick={0,0.1,0.2,0.3,0.4,0.5,0.6,0.7,0.8,0.9,1.0,1.01},
    legend pos=north west,
    ymajorgrids=true,
    grid style=dashed,
]
 
\addplot[
    color=blue,
    mark=square,
    ]
    coordinates {
    (30,0.087979) (60,0.265287) (90,0.450649) (120, 0.610879) (150, 0.732862) (180, 0.821445)
};
\addplot[
    color=magenta,
    mark=square,
    ]
    coordinates {
    (30,0.00358216) (60,0.146213) (90, 0.2793) (120,0.4072) (150, 0.5396) (180, 0.6079)
};
    
    \legend{Prob. before \textit{design principles}, Prob. after \textit{design principles}}
\end{axis}
\end{tikzpicture}}

%% file: Four_Diagrams/Privacy_Combined.tex
\centering
\resizebox{\linewidth}{!}{      
\begin{tikzpicture}
\begin{axis}[
    xlabel={Time (s)},
    ylabel={Probability of privacy leakage},        width=10cm, height=5.4cm,
    xmin=30, xmax=180,
    ymin=0, ymax=1.01,
    xtick={30, 60, 90 ,120, 150 ,180},
    ytick={0,0.1,0.2,0.3,0.4,0.5,0.6,0.7,0.8,0.9,1.0,1.01},
    legend pos=north west,
    ymajorgrids=true,
    grid style=dashed,
]
 
\addplot[
    color=blue,
    mark=square,
    ]
    coordinates {
    (30,0.0297068) (60,0.100165) (90,0.194637) (120, 0.299908) (150, 0.401481) (180, 0.499895)
};
\addplot[
    color=magenta,
    mark=square,
    ]
    coordinates {
    (30, 0.00475419) (60, 0.0155736) (90, 0.0314485) (120, 0.0523964) (150, 0.0769988) (180, 0.104888)
};
    
    \legend{Prob. before \textit{design principles}, Prob. after \textit{design principles}}
    \end{axis}
\end{tikzpicture}}       